\documentclass[conference]{IEEEtran}
\IEEEoverridecommandlockouts
\usepackage{cite}
\usepackage{amsmath,amssymb,amsfonts}
\usepackage{algorithmic}
\usepackage{graphicx}
\usepackage{textcomp}
\usepackage{xcolor}
\usepackage{enumitem}
\usepackage{enumerate}
\usepackage{subcaption}
\usepackage{listings,jvlisting}
\usepackage{inconsolata}

\usepackage{tikz} 
\usepackage{lipsum} 
\makeatletter 

\def\ieeecopyright{ \footnotesize © 2024 IEEE. Personal use of this material is permitted.\newline DOI: 10.1109/ISORC61049.2024.10551370} \makeatother \AddToHook{shipout/firstpage}{
\begin{tikzpicture}[remember picture,overlay] \node[anchor=south west,xshift=1.0cm,yshift=0.8cm] at (current page.south west){\parbox{\linewidth}{\raggedright\ieeecopyright}}; \end{tikzpicture} }

\def\BibTeX{{\rm B\kern-.05em{\sc i\kern-.025em b}\kern-.08em
    T\kern-.1667em\lower.7ex\hbox{E}\kern-.125emX}}
\begin{document}

\title{TECS/Rust: Memory-safe Component Framework\\ for Embedded Systems\\
}

\author{\IEEEauthorblockN{Nao Yoshimura}
\IEEEauthorblockA{\textit{Graduate School of}\\
\textit{Science and Engineering}\\
\textit{Saitama University}\\
}
\and
\IEEEauthorblockN{Hiroshi Oyama}
\IEEEauthorblockA{\textit{OKUMA Corporation}}
\and
\IEEEauthorblockN{Takuya Azumi}
\IEEEauthorblockA{\textit{Graduate School of}\\
\textit{Science and Engineering}\\
\textit{Saitama University}\\
}

}

\maketitle

\begin{abstract}

As embedded systems grow in complexity and scale due to increased functional diversity, component-based development (CBD) emerges as a solution to streamline their architecture and enhance functionality reuse. CBD typically utilizes the C programming language for its direct hardware access and low-level operations, despite its susceptibility to memory-related issues. To address these concerns, this paper proposes TECS/Rust, a Rust-based framework specifically designed for TECS, which is a component framework for embedded systems. It leverages Rust's compile-time memory-safe features, such as lifetime and borrowing, to mitigate memory vulnerabilities common with C. The proposed framework not only ensures memory safety but also maintains the flexibility of CBD, automates Rust code generation for CBD components, and supports efficient integration with real-time operating systems. An evaluation of the amount of generated code indicates that the code generated by this paper framework accounts for a large percentage of the actual code. Compared to code developed without the proposed framework, the difference in execution time is minimal, indicating that the overhead introduced by the proposed framework is negligible.

\end{abstract}


\section{Introduction}

In recent years, embedded systems have become increasingly complex and large-scale~\cite{FMP3+TECS, LargeScaleAndComplex}, driven by the increasing diversity of functionalities implemented in embedded devices. Embedded systems are resource limited, and performance is important. As these systems grow more complex, the challenge intensifies to design, develop, and maintain them efficiently. The pressing need for scalable, efficient, and adaptable development methodologies in the face of these challenges paves the way for approaches designed to embedded systems development.\par
\textit{Component-based development} (CBD)~\cite{TECS_ISORC2010, ThreeLayerCBD} is one of the methods to solve the increasing complexity and scale of embedded systems. CBD divides a system into components and subsystems according to their elements and functionalities. CBD facilitates a more structured and comprehensible system architecture. This modular approach assigns distinct functionalities to individual components, thereby enhancing the system's modularity and the independent reusability of each functionality~\cite{GECCO}. CBD not only simplifies the understanding of the system's structure but also significantly boosts efficiency in development processes. Most of the languages adopted for CBD in embedded systems development are C and C++ due to their well-suited compatibility with the complex requirements and performance needs inherent to these systems.\par

C and C++ are pivotal for embedded systems, particularly for devices requiring real-time safety performance. They provide direct hardware access and support essential low-level operations for superior embedded system performance. The extensive tools and libraries available enhance development efficiency and allow code reuse. However, C and C++ lack memory safety~\cite{galeed}, a critical issue for resource-constrained embedded systems that demand precise memory management. This deficiency can lead to incorrect memory references, leaks, and dangling pointers, making the languages prone to errors and vulnerabilities~\cite{benefitDrawbacks}. Therefore, choosing a programming language for secure embedded systems is crucial, underscoring the need for alternatives that address security concerns.\par

Rust emerges as a programming language that ensures memory safety at the language level~\cite{rustsdk}. Rust features a \textit{borrowing} and \textit{ownership} model, \textit{lifetime} annotations, and the elimination of null pointers. These features are rigorously enforced by the Rust compiler, which meticulously checks for potential bugs, such as null and dangling pointers, thereby preventing unintentional memory access errors. This strict compiler oversight ensures a more secure software development process. Moreover, Rust offers the same level of precise control over memory and resources as C and C++, including the capability for memory-mapped I/O, which enables direct manipulation of hardware devices. This feature is particularly advantageous in embedded systems development, where such direct control is essential for interacting with peripherals and optimizing system performance. The adoption of Rust is a necessary element for developing embedded systems with both safety and performance.\par

This paper proposes a CBD framework for embedded system development using Rust. The CBD framework for embedded systems focuses on \textit{TOPPERS Embedded Component Systems} (TECS)~\cite{TECS}. TECS is a component framework for embedded systems and is also used for \textit{real-time operating systems} (RTOSs). RTOSs, critical for achieving multitasking control and low latency in embedded environments, benefit from the integration of Rust, improving reliability and performance.

The proposed framework facilitates CBD with Rust, offering an approach to embedded system development. By incorporating Rust, the proposed framework ensures memory-safe development, while CBD principles enhance flexibility, reusability, and maintainability of the systems developed. Furthermore, the framework simplifies the use of Rust in the RTOS and enables memory-safe application development.\par
The contributions of this paper are as follows:
 \begin{itemize}
\item Implementation of CBD using Rust with minimal overhead: Unlike traditional CBD frameworks designed primarily for C, this Rust-based approach provides enhanced memory safety with negligible performance impact.
\item Efficient CBD with automatic Rust code generation: TECS automatically generates C code from dedicated component description files. Automatic Rust generation is achieved by implementing plugins that generate Rust code, making CBD more efficient.
\item Realization of low overhead operation on the RTOS: The RTOS for embedded systems has real-time performance and high reliability. A flow for making generated Rust code available to the RTOS is proposed. The framework in this paper enables Rust to run on the RTOS with minimal additional overhead.
 \end{itemize}

 This paper is organized as follows. Section II discusses the system model and the assumptions of the embedded component system in this study. Section III describes the development flow, design, and implementation. Section IV presents the evaluation results. Section V introduces and compares related studies. Section VI provides the conclusion of this study.
 
\section{System Model}

\begin{figure}[t]
\begin{center}
        \includegraphics[width=0.87\linewidth]{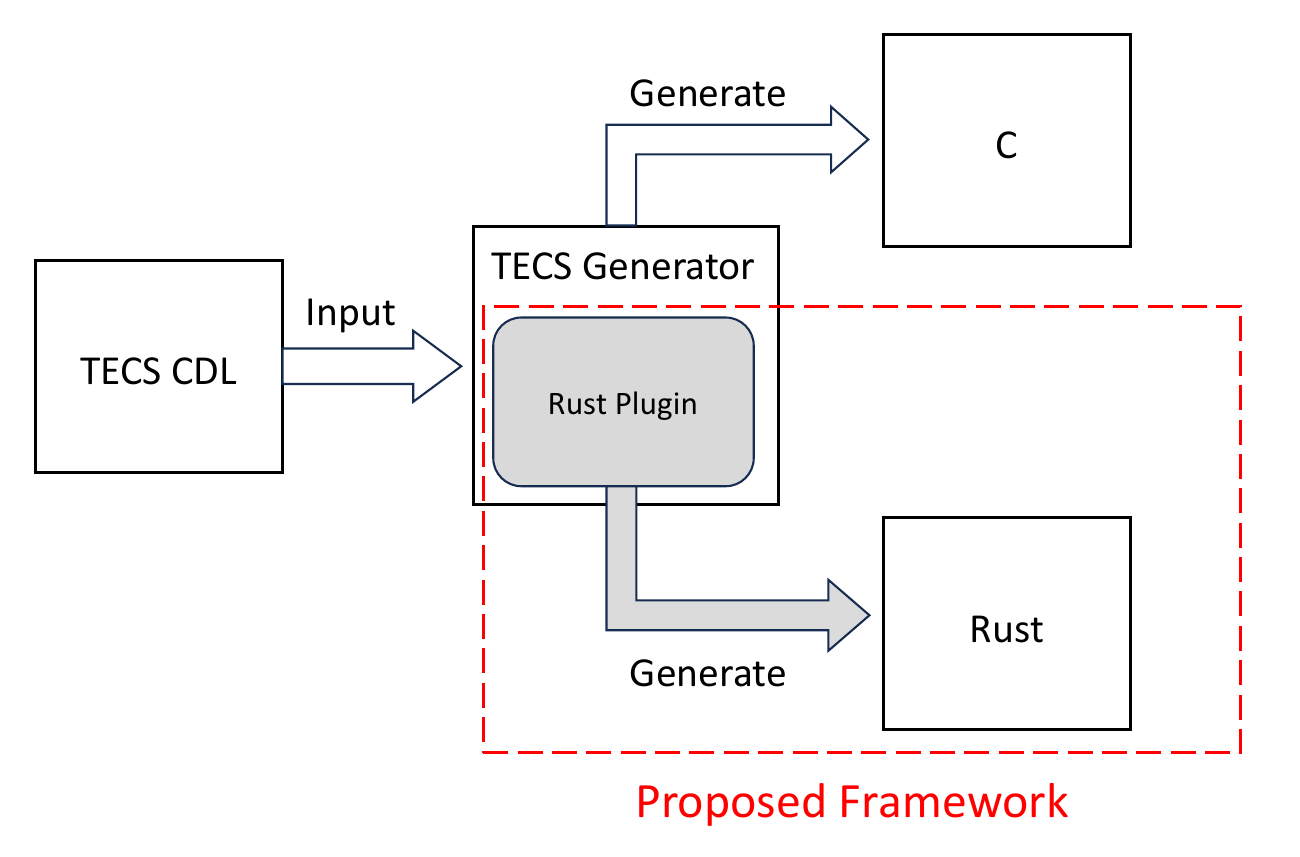}
    \vspace{-8pt}
	\caption{System model.}
    \label{fig:System model}
    \vspace{-23pt}
\end{center}
\end{figure}

This section describes the system model of the proposed framework, TECS, and Rust. The system model structure of the proposed framework is shown in Fig.~\ref{fig:System model}. Originally, the TECS generator generates files written in C by taking the component description file as input. The proposed framework creates TECS generator plugins. These plugins generate code written in Rust from the component description files. This means that Rust code is automatically generated according to the component description file as input.

\vspace{-6pt}
\subsection{TECS}

TECS is an embedded component system. TECS is a framework that supports the development of embedded software by dividing embedded software into components and subsystems. Partitioning software into components improves reusability and facilitates the reuse of components. TECS is intended for embedded software because TECS statically generates and combines components. This reduces runtime overhead and is suitable for embedded systems that require real-time performance. Componentization separates the interface from the implementation. The interface in TECS is the entity that connects components to components. This interface makes the implementation possible to change without changing the name of the calling function. This improves the maintainability of the system.
 
  \subsubsection{TECS CDL} In TECS, the definition of components and their coupling relationships are described in \textit{Component Description Language} (CDL). The main contents of the CDL description are the \textit{signature} description, the \textit{celltype} description, and the \textit{cell} description.

    \quad a) \textit{Signature} Description: The \textit{signature} description defines a \textit{signature}, which is a set of functions that interface between components. An example of a simple \textit{signature} description is shown in Fig.~\ref{fig: Signature description}. \textit{Signature} names are defined by the keyword ``\textit{signature}'' followed by ``s,'' which generally appears at the beginning of the \textit{signature} name. A function is defined by a return type, a function name, and function arguments. Each argument must have a specifier. ``[in]'' is a specifier for input, where the caller component has \textit{ownership} of the data. ``[out]'' is an output specifier that allows the destination component to change the data.
    
\begin{figure}[t]
\centering
	\begin{lstlisting}
 signature sSensor {
   void set_device_ref( void );
   void get_distance( [out] int32_t* distance );
   void light_on( void );
   void light_set( [in] int32_t bv1, [in] int32_t bv2, [in] int32_t bv3, [in] int32_t bv4 );
   void light_off( void );
 };
	\end{lstlisting}
\vspace{-7pt}
\caption{Signature description.}
\vspace{-7pt}
\label{fig: Signature description}
\end{figure}
    
    \quad b) \textit{Celltype} Description: The \textit{celltype} description defines the type of the component. The definition includes \textit{entry ports}, \textit{call ports}, attributes, and variables. A simple example of \textit{celltype} description is shown in Fig.~\ref{fig: Celltype description}. The \textit{celltype} name follows the keyword ``celltype'' and is generally prefixed with a ``t.'' The \textit{entry port} and \textit{call port} are then defined. The \textit{entry port} is a port that provides  functionalities, and the \textit{call port} is a port for calling the \textit{entry port} function. 
    These port definitions include the connecting \textit{signature}. Attributes are defined and initialized in ``\{\};'' following the keyword ``attr.'' They are fixed values and cannot be rewritten at runtime. Variables are defined in ``\{\};'' following the keyword ``var'' and their values can be changed at runtime.

\begin{figure}[t]
\centering
	\begin{lstlisting}
 [generate (RustGenPlugin, "lib")]
 celltype tSensor {
   call sPowerdown cPowerdown;
   entry sSensor eSensor;
   attr{
     pbio_port_id_t port = C_EXP("pbio_port_id_t::PBIO_PORT_ID_$port$");
   };
   var {
     Option_Ref_a_mut__pup_device_t__ ult = C_EXP("None");
   };
 };
	\end{lstlisting}
\vspace{-7pt}
\caption{Celltype description.}
\vspace{-15pt}
\label{fig: Celltype description}
\end{figure}
    
    \quad C) \textit{Cell} Description: The \textit{cell} description defines component instances, \textit{cells}, 
    and the coupling relationship between components.  Variables can be initialized with this description, and attributes can also be initialized with this description. \textit{Cell} names are defined by the keyword ``cell'' followed by a \textit{celltype} name. The binding relation of \textit{cells} is defined by joining the \textit{call port} of the \textit{cell} with the \textit{entry port} of another \textit{cell}, as shown in Fig.~\ref{fig: Cell description}.
    
\begin{figure}[t]
\centering
	\begin{lstlisting}
 [generate (RustGenPlugin, "lib")]
 cell tSensor Sensor {
   cPowerdown = Powerdown.ePowerdown2;
   port = C_EXP("pbio_port_id_t::PBIO_PORT_ID_B");
 };
	\end{lstlisting}
\vspace{-7pt}
\caption{Cell description.}
\vspace{-13pt}
\label{fig: Cell description}
\end{figure}

\begin{figure}[t]
\begin{center} 
        \includegraphics[width=1\linewidth]{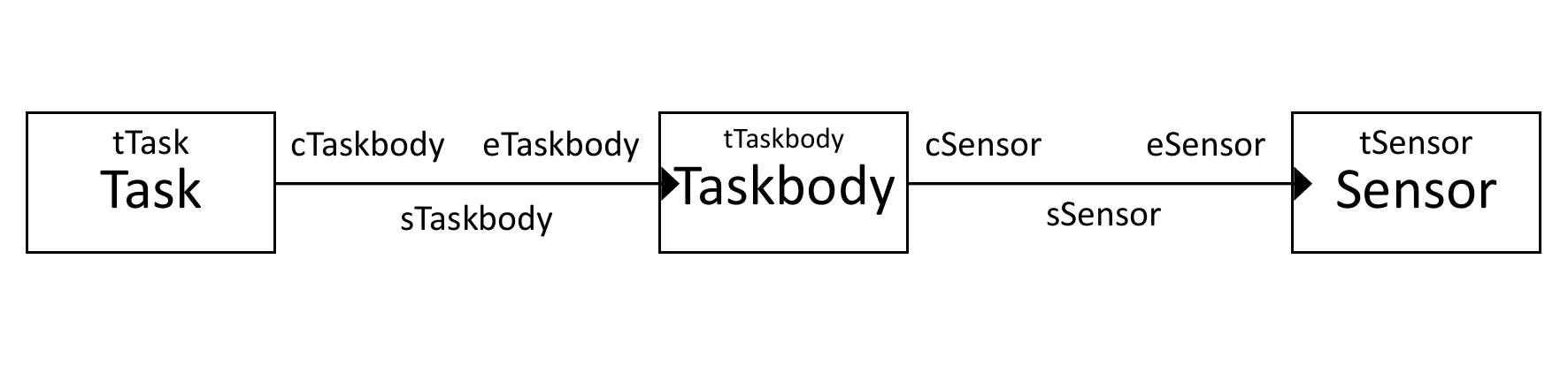}
    \vspace{-17pt}
	\caption{Component diagram.}
    \label{fig:Component diagram}
    \vspace{-23pt}
\end{center}
\end{figure}

  \subsubsection{Component Diagram} A component diagram is a graphical representation of the relationship between components. A component diagram helps to understand the structure of the system and makes development more efficient. An example of a simple component diagram is shown in Fig.~\ref{fig:Component diagram}. Each rectangle is called a \textit{cell}, which represents an instance of a component. Each \textit{cell} is marked with the \textit{celltype} name and the \textit{cell} name inside. A straight line connecting two \textit{cells} represents a \textit{signature}, which is given a \textit{signature} name. At both ends of the \textit{signature}, a \textit{call port} and an \textit{entry port} exist for binding. The \textit{entry port} is represented by a black triangle.

\vspace{-6pt}
\subsection{Rust}

Rust~\cite{Rust} is designed with a focus on memory safety, parallelism, and speed. The \textit{ownership} system uses a variable called owner, and each value in Rust corresponds to this variable, which is always a single value. In a \textit{borrowing} system, obtaining a reference to a variable is called \textit{borrowing}. By specifying a \textit{lifetime} for the \textit{borrowing}, the compiler is told how long the \textit{borrowing} is valid over what scope. This allows the compiler to verify that the \textit{borrowing} is safe. These concepts allow the Rust compiler to point out memory-related bugs such as incorrect memory references and dangling pointers.

    \subsubsection{Trait} Trait is a mechanism for defining a set of methods and enforcing the implementation of those methods on a specific type. As shown in Fig.~\ref{fig: Trait}, the keyword ``impl'' can be used to implement traits on a specific type, and functions can be implemented.

\begin{figure}[t]
\centering
	\begin{lstlisting}
 // Definition of trait.
 trait Printable {
   fn print(&self);
 }

 struct Person {
   name: String,
 }
 // Implement Printable trait in Person structure.
 impl Printable for Person {
   fn print(&self) {
     println!("Person: {}", self.name);
   }
 }
	\end{lstlisting}
\vspace{-9pt}
\caption{Trait.}
\label{fig: Trait}
\vspace{-10pt}
\end{figure}

\begin{figure}[t]
\centering
	\begin{lstlisting}
 // Generic structure.
 struct Container
 where
   T: Printable,    // Trait bound.
 {
   item: T,
 }
 // Implement methods related to generic structures.
 impl<T: Printable> Container<T> {
   fn new(item: T) -> Self {
     Container { item }
   }
   fn print_item(&self) {
     self.item.print();
   }
 }
	\end{lstlisting}
\vspace{-9pt}
\caption{Trait bound.}
\label{fig: Trait bound}
\vspace{-15pt}
\end{figure}
        
    \subsubsection{Trait Bound} Trait bound is a functionality that specifies a trait for generics and guarantees that the type implements the specified trait. A simple example of a generic structure is shown in Fig.~\ref{fig: Trait bound}, in which the generic type ``T'' is followed by a trait to use trait bound. When defining this structure, the type that implements the specified trait must be passed.

    \subsubsection{Unsafe Block} An unsafe block is a block of code that temporarily disables safety guarantees of Rust. Normally, the compiler inspects the code to ensure safety, but some operations do not obey the safety rules and require the code to be written using an unsafe block. In Rust, static mutable variables must be handled in an unsafe block. The raw pointer manipulation is necessary for the development of embedded systems. This is different from normal references and must be performed within the unsafe block. In addition, the C language, which has abundant resources, is often used from Rust. When using C from Rust, C must be used in an unsafe block. The unsafe block is a powerful feature, and incorrect use of the block can threaten memory safety. Therefore, the code in the unsafe block must be guaranteed to be memory safety by the programmer.

\vspace{-2pt}
\subsection{TOPPERS/ASP3}
TOPPERS/ASP3~\cite{ASP3} is the RTOS for embedded systems. The specification of this OS is based on ITRON, the RTOS specification. ASP3 is a standard profile in the TOPPERS kernel and does not support multiprocessor or dynamic generation of kernel objects. Dynamic memory utilization can cause memory shortages during system operation. This memory shortage is a difficult problem to deal with in embedded systems. To avoid this problem, ASP3 statically creates kernel objects. This eliminates the need for dynamic memory management in the kernel. In addition, the kernel is highly reliable, secure, and real-time.

    \subsubsection{Itron Crate} Itron crate~\cite{itroncrate} is one of the packages in Rust for using ITRON-specification OS. This crate allows the use of TOPPERS/ASP3 from Rust. Itron crate is realized by wrapping ID of an ASP3 kernel object in an opaque wrapper. This wrapper is created by calling the create method or by converting from the raw object ID. For safety, deleting objects or reusing their IDs by the wrapper should be done with caution. The ID wrappers are implemented with the ASP3 APIs, enabling the APIs to be used by creating a wrapper.
    
\section{Proposed Framework}

\begin{figure}[t]
\begin{center} 
        \includegraphics[width=1\linewidth]{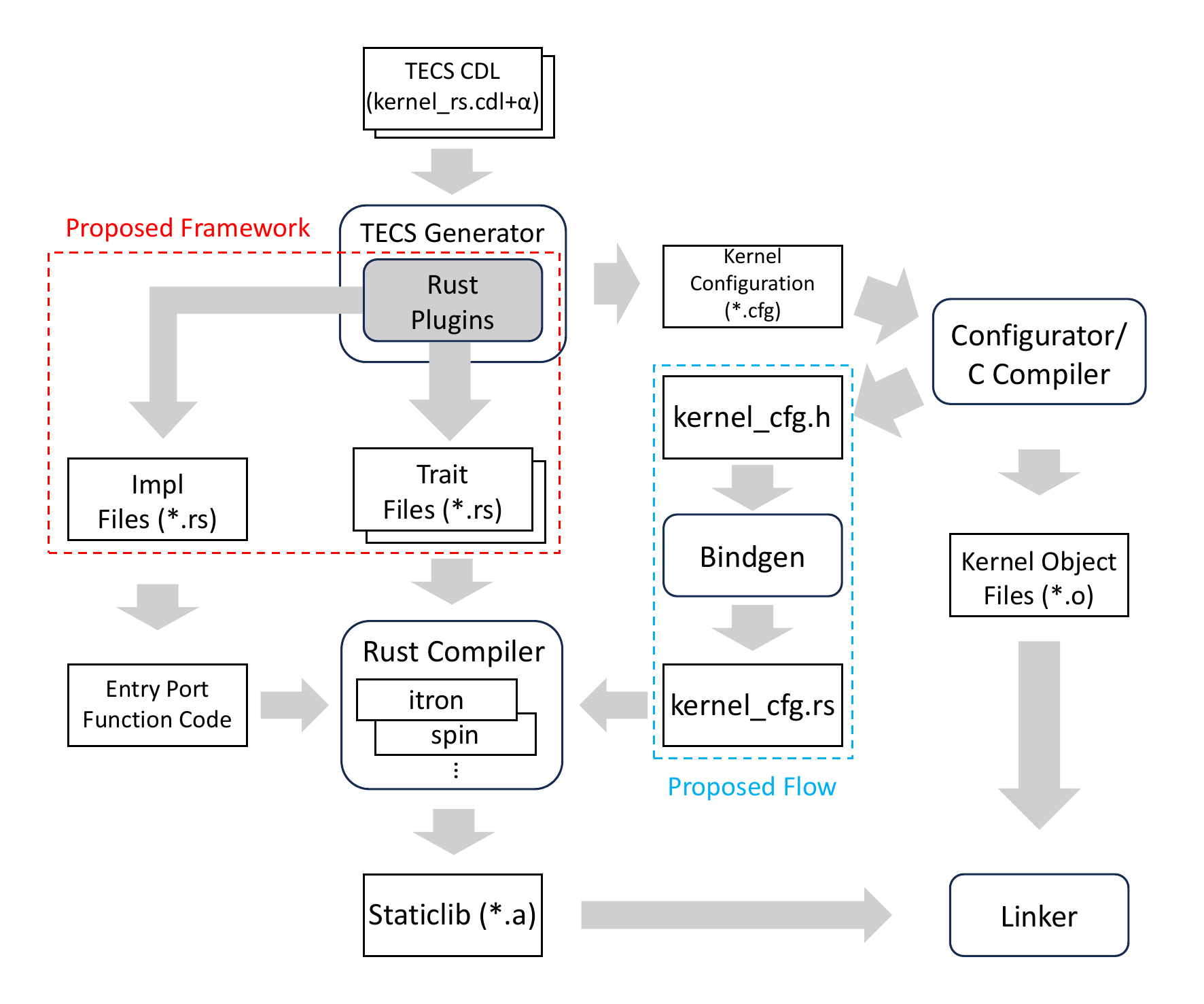}
    \vspace{-7pt}
	\caption{Flow of using Rust Plugins with ASP3.}
    \label{fig: RustPlunigFlowFigWithAsp3}
    \vspace{-19pt}
\end{center}
\end{figure}

This section describes the proposed framework, as shown in Fig.~\ref{fig: RustPlunigFlowFigWithAsp3}. The integration of a Rust plugin with the TECS generator, a key feature of the proposed framework, facilitates the conversion of TECS CDL files into corresponding Rust code. The CDL files are specified by the component specification developer and contain detailed specifications, which are manifested in the \textit{signature} and the \textit{celltype} descriptions. The \textit{cell} description in the CDL file is designed by the application developer and defines the composition of the components. The Rust plugin utilizes these CDL files to generate the corresponding Rust code. For the generation, the Rust code corresponding to each description in the CDL file must be determined. Once the code generation process is complete, the code is compiled into an executable file. This section is organized as follows: Section~\ref{ssub: Rust code for CDL files} describes the details of the Rust code corresponding to each CDL file. Section~\ref{ssub:Running the TECS generator with the Rust plugin} describes how to run the TECS generator using the Rust plugin. Section~\ref{ssub: Rust Plugin Usage Flow for RTOS} describes the flow of using the generated code in TOPPERS/ASP3.

\vspace{-4pt}
\subsection{Rust Code for CDL Files}
\label{ssub: Rust code for CDL files}

This section describes the Rust code corresponding to the CDL file. A CDL file consists of the \textit{signature} description, the \textit{celltype} description, and the \textit{cell} description. For the automatic generation of Rust code from a CDL file, the corresponding code is determined for each description. The contents of CDL file described are part of a robot application sample. The sample uses an ultrasonic sensor to monitor the distance and move the motor depending on the distance. The ultrasonic sensor component is described in this section.

\begin{figure}[t]
\centering
	\begin{lstlisting}
 pub trait SSensor {
   fn set_device_ref(&self);
   fn get_distance(&self, distance: &mut i32);
   fn light_on(&self);
   fn light_set(&self, bv1: &i32, bv2: &i32, bv3: &i32, bv4: &i32);
   fn light_off(&self);
 }
	\end{lstlisting}
\vspace{-7pt}
\caption{s\_sensor.rs.}
\label{fig: s_sensor.rs}
\vspace{-16pt}
\end{figure}

    \subsubsection{Rust Code for the \textit{Signature} Description} This section describes an example of turning the \textit{signature} description into Rust code. \textit{Signatures} in TECS are similar to traits in Rust. A \textit{signature} is a set of functions that interface between components, while a trait defines a set of methods. Therefore, traits are useful for expressing the \textit{signature} descriptions.\par
    
    The \textit{signature} description to be Rust coded is described below. The \textit{signature} shown in Fig.~\ref{fig: Signature description} is named sSensor and connects to the ultrasonic sensor component. The functions included in this \textit{signature} are set\_device\_ref, get\_distance, light\_on, light\_set, and light\_off. The light\_set function takes int32\_t type \textit{bv1}, \textit{bv2}, \textit{bv3}, and \textit{bv4} as arguments, and the return value is void. \textit{Bv1} is the argument that stores the brightness of the first light. The specifier of \textit{bv1} is ``[in],'' which makes \textit{bv1} immutable, and the caller of the light\_set function has ownership. The get\_distance function takes an argument of int32\_t type \textit{distance}, and the return value is void. \textit{Distance} has the specifier ``[out]'' and is mutable. In the CDL file, the argument with the ``[out]'' specifier must be a pointer. This is because the caller has ownership and is mutable.\par
    
    Trait is used in Rust code for the \textit{signature} descriptions. The Rust code in Fig.~\ref{fig: s_sensor.rs} corresponds to the \textit{signature} description in Fig.~\ref{fig: Signature description}, and the trait is defined. This \textit{signature} and this trait have five functions. The number of \textit{signature} functions is the same as the number of functions in the corresponding trait. The name of Fig.~\ref{fig: s_sensor.rs} trait is SSensor, which corresponds to the name sSensor in Fig.~\ref{fig: Signature description}. This follows the naming convention for traits in Rust. This trait has the functions set\_device\_ref, get\_distance, light\_on, light\_set, and light\_off. The expressions of the functions are almost the same as those of the \textit{signature} description, but the argument description is different.\par
    
    \textit{Signature} function argument correspondence is described below. In the \textit{signature} description, each argument requires a specifier. However, in Rust, these specifiers do not exist. Therefore, the specifier is corresponded to by using  ``\&'' of Rust. ``\&'' is used to indicate \textit{borrowing} in Rust. In Rust, when a function argument is a borrowed value, the argument type is prefixed with ``\&.'' 
    In Fig.~\ref{fig: s_sensor.rs}, the argument of the light\_set function, \textit{bv1}, has the type \&i32. The \&i32 indicates an immutable \textit{borrowing} of a value of type i32, and the \textit{ownership} is held by the caller of the function. The argument of the get\_distance function, \textit{distance}, has the type \&mut i32. \&mut i32 indicates mutable \textit{borrowing} of a value, and the \textit{ownership} is the same as for \&i32.

    \subsubsection{Rust Code for the \textit{Celltype} Description and the \textit{Cell} Description} This section describes the Rust code corresponding to the \textit{celltype} description and the \textit{cell} description. The \textit{celltype} description defines the type of a component and includes definitions of \textit{call port}, \textit{entry port}, attributes, and variables. In Rust code, these definitions are expressed using structures. The \textit{cell} description defines the actual instances of \textit{cells} and connection relations based on the component type definitions. In Rust code, these instantiations are static.\par

    The \textit{celltype} description to be Rust coded is described below. The \textit{celltype} in Fig.~\ref{fig: Celltype description} defines the type of the ultrasonic sensor component. The \textit{celltype} name is tSensor, which has an \textit{entry port} named eSensor and a \textit{call port} named cPowerdown. eSensor is connected by a \textit{signature} named sSensor. On the other hand, cPowerdown is connected by a \textit{signature} named sPowerdown. The \textit{port} is an attribute with type pbio\_port\_id\_t. In addition, \textit{ult} is defined as a variable. The type of \textit{ult} is Option\_Ref\_a\_mut\_\_pup\_device\_t\_\_ and \textit{ult} is initialized by C\_EXP. C\_EXP takes a string literal argument, and the argument is output as the initialization value. Based on these definitions, the cell description is created.\par

    The \textit{cell} description to be Rust coded is described below. The \textit{cell} in Fig.~\ref{fig: Cell description} is named Sensor of type tSensor. The \textit{call port} of this \textit{cell}, cPowerdown, is connected to the \textit{entry port} of the \textit{cell} named Powerdown, ePowerdown2. This \textit{call port} represents the use of Powerdown \textit{cell} functionality. The \textit{port} attribute is initialized using the C\_EXP initializer. This initialization indicates which port the ultrasonic sensor is connected to.\par
    
\begin{figure}[t]
\centering
	\begin{lstlisting}
 use spin::Mutex;
 use crate::{s_powerdown::*, t_powerdown::*, s_sensor::*};
 
 pub struct TSensor<'a, T>
 where
   T: SPowerdown,
 {
   pub c_powerdown: &'a T,
   pub port: pbio_port_id_t,
   pub variable: &'a Mutex<TSensorVar<'a>>,
 }
 
 pub struct TSensorVar<'a>{
   pub ult: Option<&'a mut pup_ultrasonic_sensor_t>,
 }
 
 pub struct ESensorForTSensor<'a>{
   pub cell: &'a TSensor<'a, EPowerdown2ForTPowerdown<'a>>,
 }
 
 pub static SENSOR: TSensor<EPowerdown2ForTPowerdown> = TSensor {
   c_powerdown: &EPOWERDOWN2FORPOWERDOWN,
   port: pbio_port_id_t::PBIO_PORT_ID_B,
   variable: &SENSORVAR,
 };
 
 pub static SENSORVAR: Mutex<TSensorVar> = Mutex::new(TSensorVar {
   ult: None,
 });
 
 pub static ESENSORFORSENSOR: ESensorForTSensor = ESensorForTSensor {
   cell: &SENSOR,
 };
 
 impl<'a, T: SPowerdown> TSensor<'a, T> {
   #[inline]
   pub fn get_cell_ref<'a>(&self) -> (&T, &pbio_port_id_t, &Mutex<TSensorVar<'a>>) {
     (&self.c_powerdown, &self.port, self.variable)
   }
 }
	\end{lstlisting}
\vspace{-7pt}
\caption{t\_sensor.rs.}
\label{fig: sensor.rs}
\vspace{-17pt}
\end{figure}

    The structure is used to convert \textit{celltype} and the \textit{cell} descriptions into Rust code. The Rust code in Fig.~\ref{fig: sensor.rs} corresponds to the \textit{celltype} description in Fig.~\ref{fig: Celltype description} and the \textit{cell} description in Fig.~\ref{fig: Cell description}, and includes the structure definition and initialization. The structure TSensor in Fig.~\ref{fig: sensor.rs} has \textit{call port}, attributes, and variables as fields. The name of the structure is based on the \textit{celltype} name of the \textit{celltype} description and is in CamelCase. The ``\textless\textgreater'' following TSensor describes lifetime annotations and generics.\par
    
    Generics are used for \textit{call port}, which is defined as a field in a structure. ``T'' is generics, which are required when using a trait bound. The \textit{call port}, cPowerdown, corresponds to the field c\_powerdown in Sensor, which follows the snake\_case convention. The type of c\_powerdown is \&T, and the trait bound enforces the implementation of the SPowerdown trait. The trait bound makes explicit to which \textit{signature} the \textit{call port} is connected. After the \textit{call port} is defined, attributes and variables are defined. Attributes are defined corresponding to the \textit{celltype} description and have the same name and number.\par
    
    Variables are defined as separate structures. A reference to the structure of the variable is held by Sensor. The reference is a structure named TSensorVar, which is determined by the \textit{celltype} description. TSensorVar references use Mutex, which provides exclusive control. TSensorVar is initialized as a static mutable structure. In Rust, static mutable values must be used in an unsafe block. Unsafe blocks are contrary to the proposed approach in terms of memory-safe development with Rust. Using static mutable values without using unsafe blocks requires the use of an exclusive control on those values.\par
    
    Exclusive control of TSensorVar uses the \textit{spin} crate, an external crate. Variable, a field of TSensor, is implemented as a \&Mutex\textless TSensorVar\textgreater~type. The references to \textit{call port} of TSensor, attribute, and variable structures are information that is statically determined at the \textit{celltype} description stage. This information can be placed in the ROM area, reducing the amount of RAM used. The RAM area is used for variables. Therefore, TSensorVar, which is a structure for variables, and TSensor, which is a structure for static information, are separated. The actual variables are defined in the TSensorVar structure, and the name and number of variables in the \textit{celltype} description are the same.\par
    
    \textit{Entry port} is defined as a completely different structure from \textit{call port}, attributes, and variables. The name of the \textit{entry port} structure is ESensorForTSensor. The name ForTSensor is used to distinguish which ESensor of \textit{celltype}. When multiple \textit{celltypes} have the same name of \textit{entry port}, a name conflict is prevented by using ForTSensor to identify which \textit{entry port} of \textit{celltype}. \par
    
    \textit{Entry port} structure has a reference to TSensor in the field. The type of the reference is TSensor\textless \textquotesingle a, EPowerdown2ForTPowerdown\textless \textquotesingle a\textgreater\textgreater. The EPowerdown2ForTPowerdown\textless \textquotesingle a\textgreater~type is generics and specifies the type of the \textit{entry port} that connects to the \textit{call port}. If TSensor has multiple \textit{call ports}, this generic specification will be multiple. This specification makes the TSensor field c\_powerdown an EPowerdown2ForTPowerdown type. This type must implement the SPowerdown trait by means of a trait bound.\par
    
    The SSensor trait must be implemented in eSensor, the \textit{entry port} of the tSensor. This corresponds to the \textit{celltype} description, because eSensor of tSensor is connected by sSensor. eSensor uses the SSensor trait to provide the implementation of the sSensor functions. In the case of SSensor trait, SSensor trait has five functions, and the contents of these functions must be implemented. This implementation is the functionality provided by tSensor. This functionality must be provided as many times as the number of \textit{entry ports} in the \textit{celltype} description. Therefore, the number of \textit{entry port} structures is the same as the number of \textit{entry ports}.\par

\begin{figure}[t]
\centering
	\begin{lstlisting}
 use spin::Mutex;
 use crate::{t_sensor::*, s_powerdown::*, s_sensor::*};
 
 impl SSensor for ESensorForTSensor<'_>{
   #[inline]
   fn set_device_ref(&self) {
     let cell_ref = self.cell.get_cell_ref();
   }
   #[inline]
   fn get_distance(&self, distance: &mut i32) {
     let cell_ref = self.cell.get_cell_ref();
   }
   #[inline]
   fn light_on(&self) {
     let cell_ref = self.cell.get_cell_ref();
   }
   #[inline]
   fn light_set(&self, bv1: &i32, bv2: &i32, bv3: &i32, bv4: &i32) {
     let cell_ref = self.cell.get_cell_ref();
   }
   #[inline]
   fn light_off(&self) {
     let cell_ref = self.cell.get_cell_ref();
   }
 }
	\end{lstlisting}
\vspace{-7pt}
\caption{t\_sensor\_impl.rs.}
\label{fig: sensor_impl.rs}
\vspace{-17pt}
\end{figure}

    The instantiation of \textit{cells} in the Rust code is described. The Rust code in Fig.~\ref{fig: sensor.rs} includes the instantiation of a \textit{cell} and corresponds to the \textit{cell} description. The names of SENSOR, SENSORVAR, and ESENSORFORSENSOR shown in Fig.~\ref{fig: sensor.rs} are determined by the names of the \textit{cell} descriptions. If multiple \textit{cells} of type tSensor are defined, instances are created in the same number as the definitions. These instances are defined as global static variables by ``pub static.'' The names of static variables should be capitalized according to the Rust naming conventions.\par
    
    The function get\_cell\_ref, which takes \textit{cell} information as the return value is explained. This function is defined for each \textit{celltype}, such as TSensor in Fig.~\ref{fig: sensor.rs}. The return type of this function is a tuple. In the case of TSensor, the function returns a tuple that contains references to the \textit{call port}, attributes, and variables of TSensor. This tuple allows easy access to the information in TSensor when implementing functions.

\vspace{-5pt}
\subsection{Running the TECS Generator with the Rust Plugin}
\label{ssub:Running the TECS generator with the Rust plugin}

     The use of the Rust plugin is declared in the CDL file. The Rust plugin generates Rust files (*.rs) corresponding to the CDL file in the format described in Section~\ref{ssub: Rust code for CDL files}. The Rust plugin is declared as shown in Figs.~\ref{fig: Celltype description} and~\ref{fig: Cell description}, and the name of the plugin is ``RustGenPlugin.'' The Rust plugin can be used by adding ``[generate( RustGenPlugin, "lib")]'' before the \textit{celltype} or \textit{cell} description.\par 
     
     The Rust trait files are generated by using the CDL file with the declaration as input to the TECS generator, as shown in Fig.~\ref{fig: RustPlunigFlowFigWithAsp3}. The trait file contains the Rust code corresponding to the \textit{signature} description in Fig.~\ref{fig: s_sensor.rs}. Trait files are generated as many times as the number of the \textit{signature} descriptions, and the file names correspond to the names of the \textit{signatures}. If the name of the \textit{signature} is sSensor, a file is generated as shown in Fig.~\ref{fig: s_sensor.rs}, and the file is named s\_sensor.rs.\par

     Files that contains definitions and initialization are generated simultaneously with the trait files, as shown in Fig.~\ref{fig: RustPlunigFlowFigWithAsp3}. The number of files including definition and initialization is the same as the number of \textit{celltypes} defined in the \textit{celltype} description. If the name of the \textit{celltype} is tSensor, a file named t\_sensor.rs is generated. The contents of t\_sensor.rs are shown in Fig.~\ref{fig: sensor.rs}, including the Rust code corresponding to the \textit{celltype} description in Fig.~\ref{fig: Celltype description} and the \textit{cell} description in Fig.~\ref{fig: Cell description}. \par

     The impl files are generated when a \textit{celltype} with \textit{entry ports} is defined in the \textit{celltype} description. The name of the impl file is t\_sensor\_impl.rs when the \textit{cell} name is Sensor. The contents of t\_sensor\_impl.rs is shown in Fig.~\ref{fig: sensor_impl.rs} and includes the implementation of trait. The implementation of the trait is done by the component developer, and the behavior of the component is described in the impl file. The separation of the description part of the developer enhances the reusability of the impl file.\par

    \subsection{Rust Plugin Usage Flow for the RTOS}
    \label{ssub: Rust Plugin Usage Flow for RTOS}

    This section describes the flow of using the Rust code generated by the proposed approach with TOPPERS/ASP3. The Rust code generated in Section~\ref{ssub:Running the TECS generator with the Rust plugin} is not intended to be used with ASP3. Therefore, code related to itron crate and configuration files for ASP3 must be written by the developer. To use ASP3 easily, a plugin that automatically generates Rust code related to itron crate and CDL file are used. ASP3 cannot be used only with the files automatically generated by the TECS generator. This problem is solved by using an external tool called Bindgen~\cite{bindgen}.

\begin{figure}[t]
\centering
	\begin{lstlisting}
 use crate::kernel_cfg::*;
 use itron::abi::*;
 use itron::TaskRef::*;
	\end{lstlisting}
\vspace{-7pt}
\caption{Generate code for itron crate.}
\label{fig: Generate code for itron-rs}
\vspace{-8pt}
\end{figure}

\begin{figure}[t]
\centering
	\begin{lstlisting}
 let task1_ref:TaskRef = unsafe{TaskRef::from_raw_nonnull(NonZeroI32::new(TSKID_1).unwrap())};
 
 task1_ref.activate().expect("activate task1")
	\end{lstlisting}
\vspace{-7pt}
\caption{An example of the use of a TaskRef object.}
\label{fig: An example of the use of a TaskRef object}
\vspace{-11pt}
\end{figure}

\begin{figure}[t]
\centering
	\begin{lstlisting}
 pub const TNUM_TSKID: i32 = 1;
 pub const TSKID_1: i32 = 1;
 pub const TNUM_SEMID: i32 = 0;
 pub const TNUM_DTQID: i32 = 0;
 pub const TNUM_ISRID: i32 = 1;
 pub const ISRID_tISR_SIOPortTarget1_ISRInstance: i32 = 1;
 pub const TNUM_INIRTN: i32 = 2;
 pub const TNUM_TERRTN: i32 = 2;
	\end{lstlisting}
\vspace{-7pt}
\caption{An example of kernel\_cfg.rs.}
\label{fig: An example of kernel_cfg.rs}
\vspace{-14pt}
\end{figure}

    \subsubsection{Rust Plugin for the RTOS}
    The plugin for the RTOS automatically generates code related to itron crate. The plugin is named ``ItronrsGenPlugin'' and inherits from ``RustGenPlugin.'' The generated code is shown in Fig.~\ref{fig: Generate code for itron-rs}. This code includes use statements for using the itron crate and objects in kernel\_cfg.rs. The use statement for the itron crate allows the use of the TaskRef of itron crate in the file. \par
    
    The TaskRef is a wrapper for a task object, which is an ASP3 kernel object. An example of the use of a TaskRef object is shown in Fig.~\ref{fig: An example of the use of a TaskRef object}, where initialization is done first. The initialization method ``from\_raw\_nonnull()'' is unsafe and must be enclosed in an unsafe block. The last argument, TSKID\_1 is the ID of the task object. After the initialization of TaskRef, the task object ID of the argument corresponds to ``task1\_ref.'' This kernel object ID is described in kernel\_cfg.rs. An example of kernel\_cfg.rs is shown in Fig.~\ref{fig: An example of kernel_cfg.rs}. The ASP3 task object API is implemented for TaskRef. By calling ``activate()'' in Fig.~\ref{fig: An example of the use of a TaskRef object}, ``\textit{act\_tsk}()'' in ASP3 is called. In addition to this API, other ITRON specification task APIs are also implemented.

    \subsubsection{Flow of Using the RTOS from Rust}
    This section describes the flow of using a project written in Rust with TOPPERS/ASP3. The overall flow is shown in Fig.~\ref{fig: RustPlunigFlowFigWithAsp3}, using the TECS generator and the respective compilers. As an overview of the flow, the TECS generator is first used to generate files. Then, the configuration file, which is one of the generated files, becomes an input to the ASP3 configurator. The configurator and compiler then generate kernel\_cfg.h and other object files. Kernel\_cfg.h is the input to Bindgen, which generates the Rust file kernel\_cfg.rs. All Rust files generated in this flow, including Kernel\_cfg.rs, are compiled by Rust compiler. Finally, the compiled files are linked by the linker. The whole process starts with CDL files such as kernel\_rs.cdl. Therefore, the contents of kernel\_rs.cdl must be understood.\par

\begin{figure}[t]
\centering
	\begin{lstlisting}
 celltype tTask_rs {
   [inline] entry	sTask	eTask;
   [inline] entry	siTask	eiTask;
   call	sTaskBody	cTaskBody;
 
   [inline] entry	siNotificationHandler	eiActivateNotificationHandler;
   [inline] entry	siNotificationHandler	eiWakeUpNotificationHandler;
   attr {
 	[omit]ID		id = C_EXP("TSKID_$id$");
 	TaskRef			task_ref = C_EXP("unsafe{TaskRef::from_raw_nonnull(NonZeroI32::new(TSKID_$id$).unwrap())}");
 	[omit] ATR		attribute = C_EXP("TA_NULL");
 	[omit] PRI		priority;
 	[omit] size_t	stackSize;
   };
   factory {
     write("tecsgen.cfg",
     "CRE_TSK(%s, { %s, 0, task_rs, %s, %s, NULL });",
     id, attribute, priority, stackSize);
   };
   FACTORY {
     write("tecsgen.cfg", "#include \"$ct$_tecsgen.h\"");
     write("$ct$_factory.h", "#include \"kernel_cfg.h\"");
   };
 };
	\end{lstlisting}
\vspace{-7pt}
\caption{Task object in the kernel\_rs.cdl.}
\label{fig: Task object in the kernel_rs.cdl}
\vspace{-13pt}
\end{figure}

    Kernel\_rs.cdl is a componentized Rust version of the kernel object. The \textit{celltype} of the task object, one of the kernel objects, is shown in Fig.~\ref{fig: Task object in the kernel_rs.cdl}. This \textit{celltype} definition is based on the C version of kernel.cdl. The \textit{call port} and \textit{entry port} of this \textit{celltype} have the same structure as kernel.cdl.\par
    
    The attribute \textit{task\_ref} of tTask\_rs is of type TaskRef and requires the use of itron crate. Task\_Ref is initialized with the string ``NonZeroI32::new(TSKID\_\$id\$).unwrap().'' This string is the same as the initialization method of TaskRef shown in Fig.~\ref{fig: An example of the use of a TaskRef object}. The ``TSKID\_\$id\$'' part of the initialization is determined by the \textit{cell} description. This value must be the same as the initialized value of \textit{id}. This is to correspond the contents of the configuration file to \textit{cells} of type tTask\_rs.

\begin{figure}[t]
\centering
	\begin{lstlisting}
 INCLUDE("tecsgen.cfg");
 
 CRE_TSK(TSKID_1, { TA_ACT, 0, task_rs, MID_PRIORITY, STACK_SIZE, NULL });
	\end{lstlisting}
\vspace{-7pt}
\caption{An example of the use of a configuration file.}
\label{fig: An example of the use of a configuration file}
\vspace{-5pt}
\end{figure}

    The configuration file describes the creation information and initial state of kernel objects. This information is described by a static API. An example of the static API for a task object is shown in Fig.~\ref{fig: An example of the use of a configuration file}. The static API ``CRE\_TSK'' is used to generate task objects. ``CRE\_TSK'' and the configuration file are generated by the TECS generator by defining a \textit{cell} of type tTask\_rs. The attributes of the \textit{cell}, such as distinguished name of the object, are initialized in the \textit{cell} description. The developer writes a \textit{cell} of type tTask\_rs in a CDL file, and ``CRE\_TSK'' and the arguments are automatically generated.\par
    
   The configuration file with static APIs is the input for the ASP3 configurator. The configurator interprets the configuration file and generates files including kernel configuration and initialization information. These files are compiled into object files by the compiler. One of the files generated is a kernel header file.\par

\begin{figure}[t]
\centering
	\begin{lstlisting}
 #define TNUM_TSKID	1
 #define TSKID_1	1
 #define TNUM_SEMID	0
 #define TNUM_DTQID	0
 #define TNUM_ISRID	1
 #define ISRID_tISR_SIOPortTarget1_ISRInstance	1
 #define TNUM_INIRTN	2
 #define TNUM_TERRTN	2
	\end{lstlisting}
\vspace{-7pt}
\caption{An example of the use of a kernel header file.}
\label{fig: An example of the use of a kernel header file}
\vspace{-10pt}
\end{figure}

    The kernel header file contains the definitions necessary to use the kernel. An example of a kernel header file is shown in Fig.~\ref{fig: An example of the use of a kernel header file}, named kernel\_cfg.h. This file contains definitions of object ID numbers and the number of registered objects (TNUM\_*). The object ID is mapped to the identifier of the object, which is the argument of the static API. The ID of the object is required for the initialization of the TaskRef type. The object IDs in kernel\_cfg.h are in C code, and must be converted to Rust. \par
    
    Object IDs in Rust are generated using Bindegen. Bindgen is an external tool to convert C code to Rust code. The converted Rust code using Bindgen is shown in Fig.~\ref{fig: An example of kernel_cfg.rs}. The name of the file generated by Bindgen is kernel\_cfg.rs by setting ``-o kernel\_cfg.rs.'' Kernel\_cfg.rs is compiled by Rust compiler together with the Rust file generated by the TECS generator.\par

    Compiling the Rust code is the last stage of the flow process in this section. The Rust code is compiled by Rust compiler, but the dependencies must include itron crate. The compilation target must be the same as in ASP3. The compiled file is made into a static library file (*.a) to be passed to the linker. The static library file is generated by specifying ``staticlib'' to Rust compiler. Finally, the static library file and the kernel object file are made an executable file by the linker.

\section{Evaluation}
\label{ssub: Evaluation}

In this section, Rust and the proposed framework are compared, and the proposed framework is evaluated. To evaluate CBD with Rust, the execution time is measured. Then, to evaluate the automatic generation of the proposed framework, the generated code and the compiled code are compared. Finally, to evaluate the execution on the RTOS, execution times of service calls are measured. The RTOS used for the measurements is TOPPERS/ASP3. The execution time is measured by running TOPPERS/ASP3 on STM32F413VG.

\begin{figure}[t]
\begin{center}
    \begin{minipage}[t]{0.47\linewidth}
    \begin{center} 
        \includegraphics[width=1\linewidth]{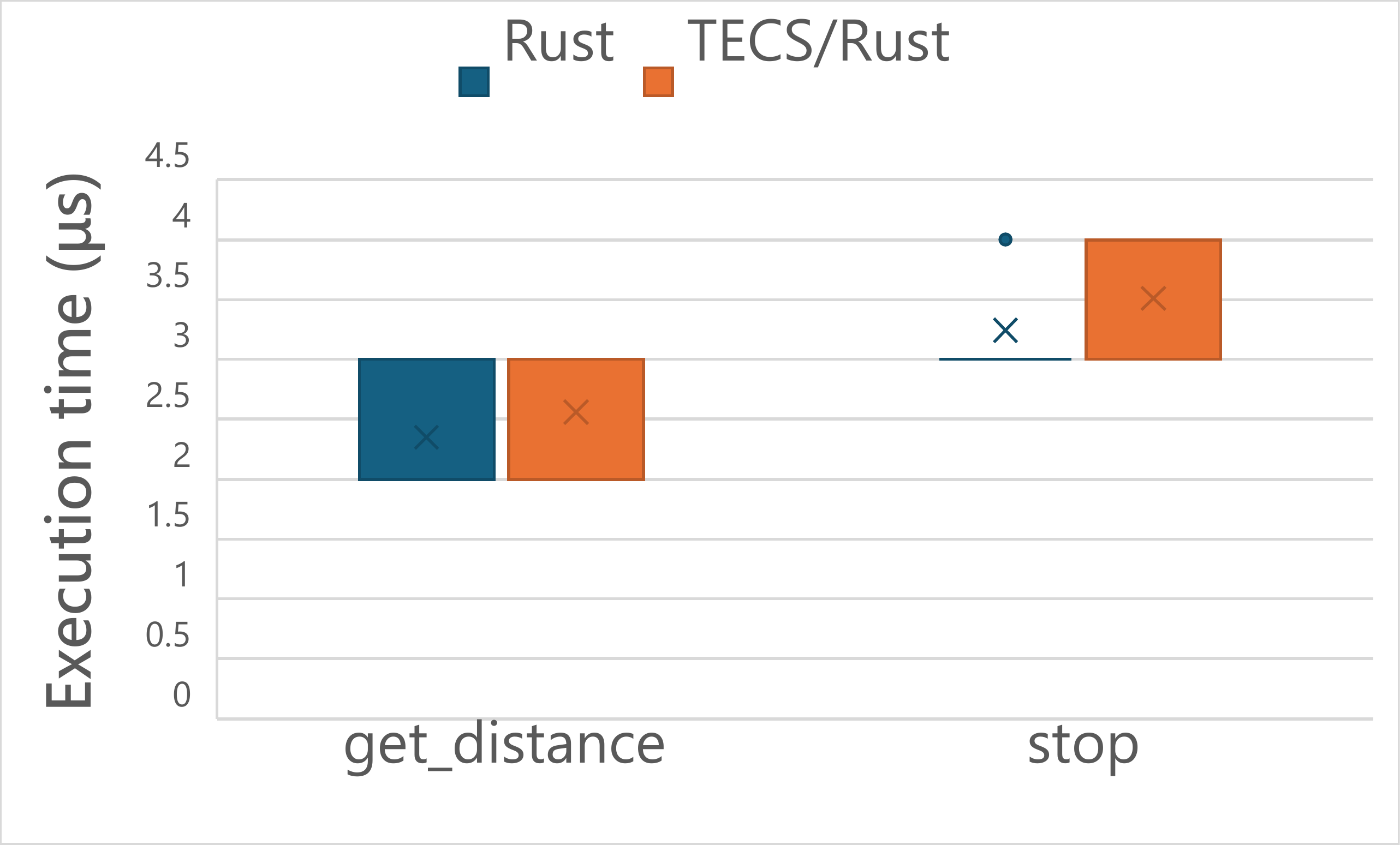}
    \end{center}
    \end{minipage}
    \begin{minipage}[t]{0.47\linewidth}
    \begin{center} 
        \includegraphics[width=1\linewidth]{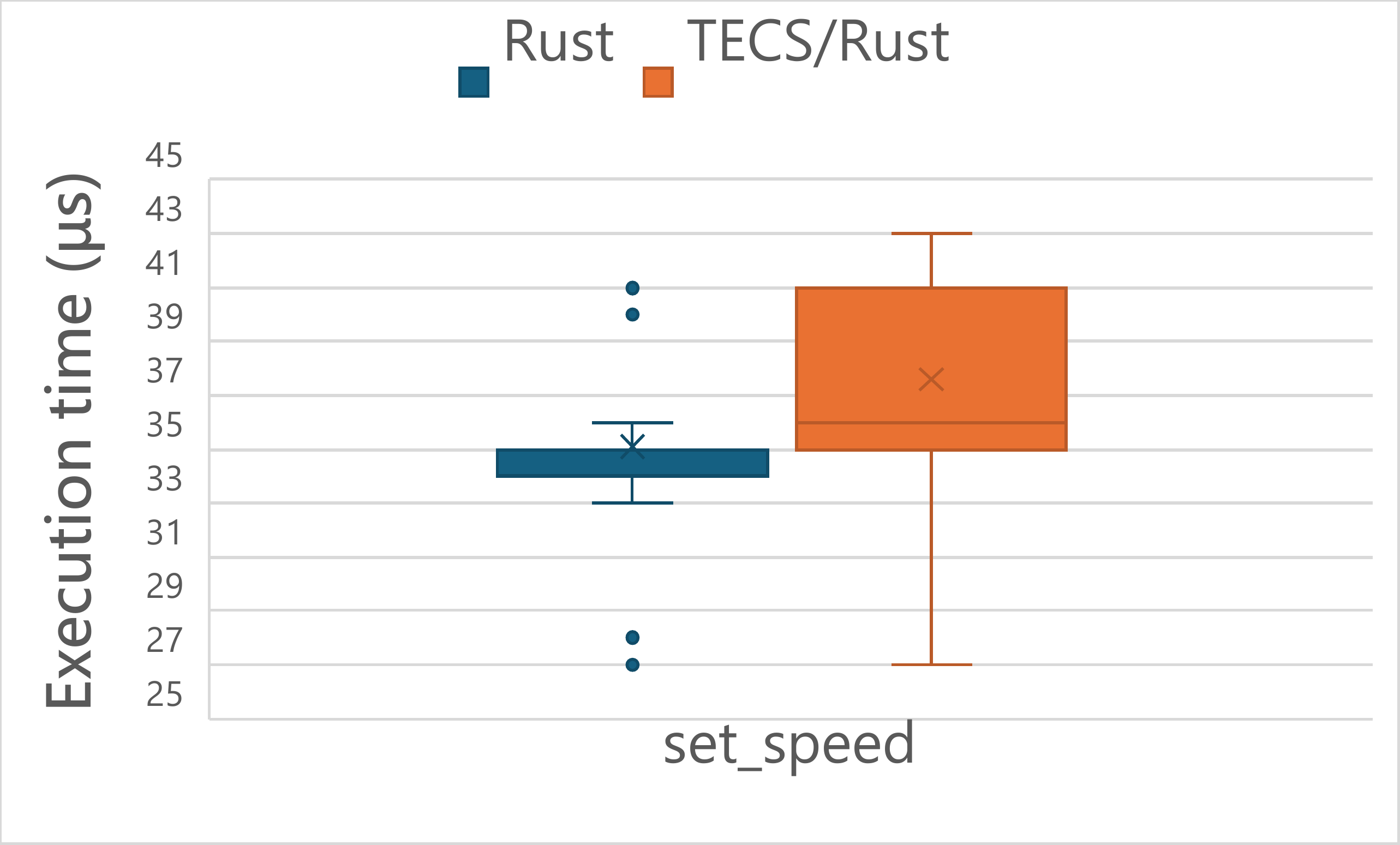}
    \end{center}
    \end{minipage}
\caption{Execution time of SPIKE-RT APIs.}
\label{fig: spikert-APIs}
\vspace{-17pt}
\end{center}
\end{figure}

\vspace{-5pt}
    \subsection{Comparison with Rust Execution Time}
    In this section, a comparison of execution times between normal Rust and Rust using TECS is presented. The comparison is performed with the APIs used in the SPIKE-RT~\cite{SPIKE-RT} sample. The three APIs used for the comparison are get\_distance, stop, and set\_speed. The comparison confirms the overhead and changes caused by the proposed approach. \par

    The execution time results are shown in Fig.~\ref{fig: spikert-APIs}, with Rust results in blue and TECS/Rust results in orange. The get\_distance is an API that retrieves the distance from an ultrasonic sensor. The comparison results show that the average of TECS/Rust is slightly slower than the average of Rust, but overall the results are almost the same. This result is similar to the result of stop, which is an API to stop a motor. On the other hand, the results for set\_speed show a larger execution time than the other APIs, making overall overhead easier to understand. This overhead is due to the exclusive control by \textit{spin} crate. The exclusive control is used in the variables of the motor and ultrasonic sensor components. The variables are used when calling the SPIKE-RT API, which causes the overhead due to the exclusive control.\par

    As a future prospect, exclusive control with low overhead needs to be implemented. The need for exclusive control is guaranteed by the component structure, which allows unsafe code in Rust. Unsafe code allows the use of static mutable variables in Rust. This reduces the overhead because the number of exclusion controls can be reduced. Therefore, the overhead caused by the exclusion control in the current results can be reduced.\par

    Common to all API results is that the TECS/Rust average is slower than the Rust average. This indicates that the overhead is caused by the componentization of TECS. However, this overhead is not large and does not have a significant impact on the system. Therefore, the proposed approach can achieve system componentization without significant overhead.

\begin{table}[t]
    \centering
    \caption{Comparison of the number of lines of code}
    \label{tab:Comparison of generated and compiled code}
    \vspace{-3pt}
    \scalebox{1}{
        \begin{tabular}{|c|c|c|c|}
            \hline
               & Rust & TECS/Rust & TECS/C \\ \hline
            CDL File & 0 & 79 & 73 \\ \hline
            Auto-generated Code & 0 & 386 & 1181 \\ \hline
            Written Code & 232 & 199 & 66 \\ \hline
            Hand-coding & 232 & 278 & 139 \\ \hline
            Compiled Code & 232 & 585 & 1247 \\ \hline
        \end{tabular}
    }
    \vspace{-5pt}
\end{table}

    \subsection{Comparison of Generated and Compiled Code}
    In this section, the number of lines of the code generated by the TECS generator and the number of lines of compiled code with hand-coding added are compared. In addition, consider the number of lines of the code in the CDL file required to use TECS. These codes are simple samples that run on SPIKE-RT, a software platform for LEGO Education SPIKE Prime. The SPIKE-RT sample is a platform that runs on TOPPERS/ASP3. The sample measures the distance to an object using an ultrasonic sensor and controls a motor according to the distance.\par
    
    The number of lines of code in the CDL file should be considered as hand-coding. CDL files are required for TECS/Rust and TECS/C, but not for Rust. The number of lines of the code in the CDL file is the difference between TECS/Rust and Rust. Therefore, the amount of hand-coding is the sum of the number of lines of the code in the CDL file and the number of lines of written code. Rust has the lowest amount of hand-coding, but this code has low reusability. On the other hand, code reusability is high for TECS/Rust and TECS/C. Although the amount of hand-coding is high for TECS/Rust and TECS/C, the CDL file makes the code highly reusable.\par

    The amount of written code for TECS/Rust shown in Table~\ref{tab:Comparison of generated and compiled code} is less than the result for Rust. The amount of written code for TECS/Rust is lower because the auto-generated code includes the code to implement functions. The automatic generation by the plugin reduces the amount of written code while maintaining a high level of code reusability. In addition, the Rust and TECS/Rust code includes the APIs of SPIKERT wrap code. This is because SPIKERT does not provide APIs for Rust. On the other hand, the APIs of SPIKERT do not need to be wrapped in C code, because SPIKERT provides C APIs. Therefore, the amount of written code in TECS/C is smaller than in TECS/Rust.\par
    
    The amount of auto-generated code by the TECS generator in TECS/Rust is less than in TECS/C. The amount of auto-generated code in TECS/Rust is approximately one-third that of TECS/C. This indicates that trait is appropriate for expressing the TECS \textit{signature}. TECS/Rust is still a temporary generation and will increase or decrease from this value in the future. At least, future optimization will reduce the amount of auto-generated code in TECS/Rust. Code generation with TECS requires only a few lines of text in the CDL file. With only a few simple declarations, TECS generates a large amount of highly reusable code. Therefore, a plugin that generates code can make CBD more efficient. In addition, CDL files can be reused in other systems, reducing the number of lines of code in future CDL files.

\vspace{-3pt}
    \subsection{Comparison of Execution Times on the RTOS}
    This section compares the execution times of Rust and TECS/Rust on TOPPERS/ASP3. The comparison was made by measuring the APIs of ASP3 task objects. To use these APIs from Rust, API wraps of itron crate were used. Itron crate is used and measured by both Rust and TECS/Rust. The comparison confirms the overhead of running the code of the proposed approach on ASP3. \par

\begin{figure}[t]
\begin{center} 
        \includegraphics[width=1\linewidth]{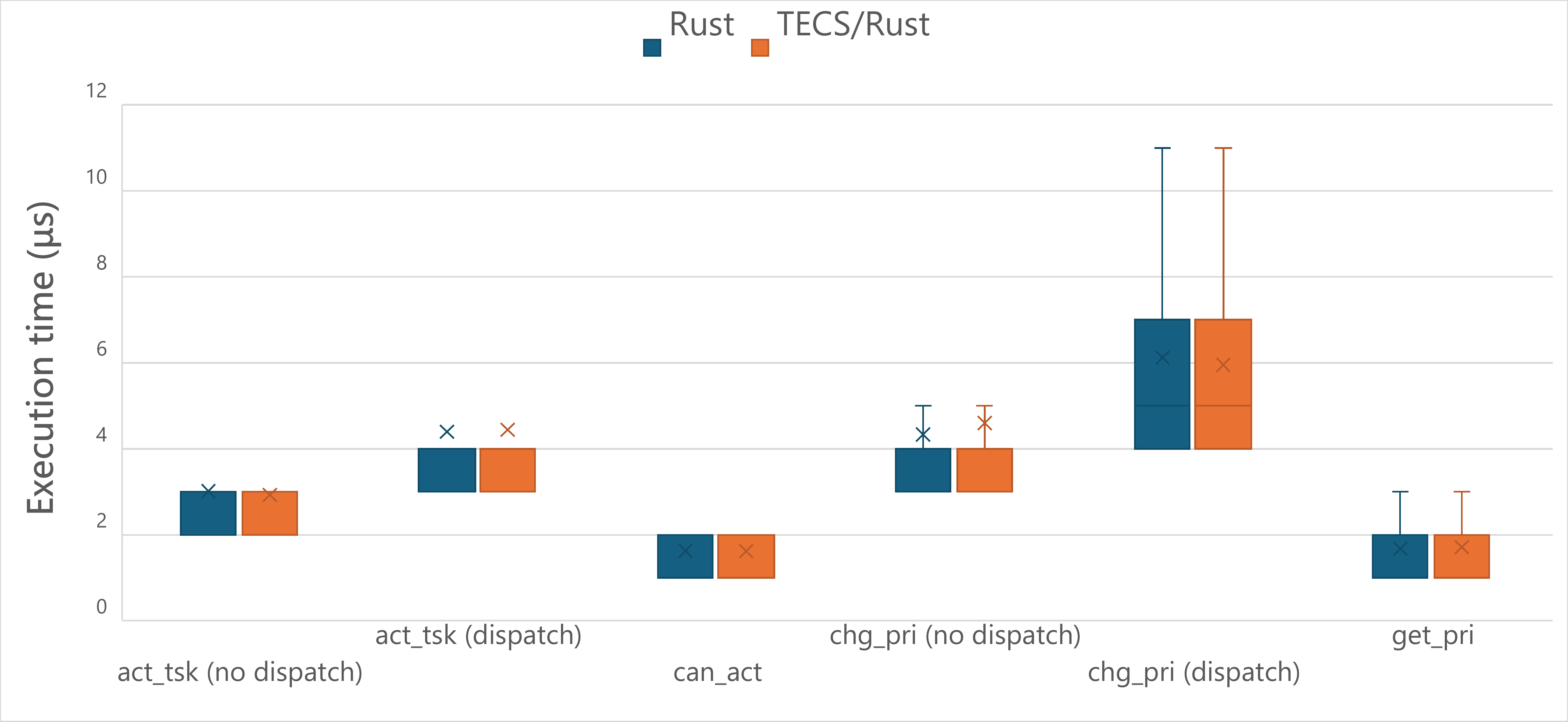}
	\caption{Execution time of task object APIs.}
    \label{fig: task_object_apis}
    \vspace{-10pt}
\end{center}
\end{figure}

    The results of measuring the task object APIs are shown in Fig.~\ref{fig: task_object_apis}. The APIs measured are \textit{act\_tsk}, \textit{can\_act}, \textit{chg\_pri}, and \textit{get\_pri}. \textit{act\_tsk} is the API for starting a task, and \textit{can\_act} is the API for canceling a task startup request. \textit{chg\_pri} is an API that changes the priority of a task, and \textit{get\_pri} is an API that obtains the priority of a task. \textit{act\_tsk} and \textit{chg\_pri} measure both cases in which dispatching occurs and does not occur. The results for all APIs show no difference between TECS/Rust and Rust, indicating that no overhead is caused by TECS. One factor in the lack of overhead is the explicit inlining of functions.\par
    
    Explicit inlining is performed in both Rust and TECS/Rust. The code generated by TECS/Rust is explicitly inlined with ``\#[inline]'' as shown in Fig.~\ref{fig: sensor_impl.rs}. The proposed approach achieves componentization with low overhead in developing Rust applications on ASP3.\par
    
    Componentization without exclusive control reduces the overhead. Exclusive controls are not used in the task components. Unlike the SPIKE-RT API, the exclusive control overhead is not incurred because no components have variables. Therefore, componentization without the use of variables results in execution time with less overhead in the proposed approach.

\vspace{-3pt}
\section{Related Work}

This section introduces existing research on Rust or CBD. The proposed framework is compared with existing research in terms of embedded system, versatility, component framework, and memory safety. The comparison between the proposed framework and existing research is shown in Table~\ref{related}.\par

\begin{table}[t]
    \caption{Comparison of the proposed framework with other methods}
    \centering
    \scalebox{0.65}{
    {\tabcolsep = 1.25mm
    \begin{tabular}{|l|c|c|c|c|c|c|}
    \hline
        & \textbf{Embedded system} & \textbf{Versatility} & \textbf{Component framework} & \textbf{Memory safety} \\
        \hline
        Tock~\cite{TockOS} & \checkmark & \checkmark & & \checkmark \\
        \hline
        CRC~\cite{CRC} & \checkmark & for concurrent program & \checkmark & \checkmark \\
        \hline
        gtk-rs~\cite{gtk-rs} & &  &  & \checkmark \\
        \hline
        AUTOSAR~\cite{AUTOSAR} & \checkmark & for automotive systems & \checkmark & \\
        \hline
        ASP3+TECS~\cite{ASP3+TECS} & \checkmark & \checkmark & \checkmark & \\
        \hline
        HRMP3+TECS~\cite{HRMP3+TECS} & \checkmark & \checkmark & \checkmark & memory protection \\
        \hline
        mruby on TECS~\cite{LightweightRuby} & \checkmark & \checkmark & \checkmark & \\
        \hline
        Dynamic CBD~\cite{DynamicCBD} & \checkmark & \checkmark & \checkmark & \\
        \hline
        Android~\cite{Android} & \checkmark & \checkmark & \checkmark & \\
        \hline
        This paper & \checkmark & \checkmark & \checkmark & \checkmark \\
        \hline
    \end{tabular}
    }
    }
    \label{related}
    \vspace{-5pt}
\end{table}

    \subsection{Memory-safe Framework Employing Rust}
    
    The framework for Tock adds a low-latency real-time component to Tock. Tock is written in Rust, a memory-safe OS for embedded systems. The framework adds low-latency functionality to the existing Tock, which lacks low-latency functionality. The proposed framework is a memory-safe framework for CBD, which is the difference from the framework for Tock.\par
    
    The framework for concurrent programming in Rust is a component framework for embedded systems. The system is modeled and analyzed by \textit{Concurrent Reactive Component} (CRC). The analysis enables preemptive execution without contention and deadlocks in the system. After analysis, Rust code is automatically generated based on the CRC. This framework is intended for concurrent programming and is not versatile. On the other hand, the proposed framework is more versatile, which is the difference from the CRC framework.\par
    
    Gtk-rs is a wrapper library for GTK+ in Rust. GTK+ is a toolkit for developing GUI applications and is used in the Linux desktop environment. In Gtk-rs, memory safety is guaranteed at the language level by using Rust. Gtk-rs is intended for the Linux desktop environment, not for embedded systems. Therefore, the difference between Gtk-rs and the proposed framework is the target of the framework.

\vspace{-1pt}
    \subsection{CBD for Embedded Systems}
    \textit{Automotive open system architecture} (AUTOSAR) is a standardized software architecture specification used in the automotive industry. AUTOSAR leverages the CBD to improve software independence and reusability. AUTOSAR is an automotive component framework and is not generic. In addition, AUTOSAR memory safety is the responsibility of the developer according to the specification. Therefore, AUTOSAR and the proposed framework differ in terms of versatility and memory safety.
    
    ASP3+TECS is one of the studies related to TECS. ASP3+TECS is a component framework for TOPPERS/ASP3, and ASP3 is componentized by TECS. The application part of this framework is developed in C using TECS. The memory safety of the application must be guaranteed by the programmer. In the proposed framework, memory safety of applications is guaranteed by the framework. Therefore, ASP3+TECS differs from the proposed framework in terms of application memory safety.\par
    
    HRMP3+TECS is a study on TECS. HRMP3+TECS is a component framework for the RTOS TOPPERS/HRMP3, which is an extension of ASP3 for multiprocessors and memory protection functionality. Memory protection is a technology to protect computer memory from unauthorized access and data corruption. The memory protection functionality of HRMP3 can be adapted to the application. Memory protection functionality prevents external attacks, and memory safety aims to prevent intrinsic memory bugs in programs. Therefore, HRMP3+TECS and the proposed framework differ in their approaches to memory.\par
    
    \textit{mruby on TECS} is one of the research projects related to TECS. mruby is a lightweight Ruby programming language, and use in embedded systems is the main focus. \textit{mruby on TECS} is a framework for using mruby with TECS. The extended framework improves the efficiency of software development in the RTOS. In addition, multiple mruby programs can be executed concurrently or in parallel, and synchronous execution is also supported. This \textit{mruby on TECS} frameworks realize CBD by mruby. Since mruby is not a memory-safe programming language, \textit{mruby on TECS} differs from the proposed memory-safe frameworks.\par
    
    A dynamic CBD framework is studied in \textit{mruby on TECS}. This framework achieves dynamic exchange of components during application execution. Dynamic exchange is difficult for components written in C because they must be compiled. For this reason, mruby is used to achieve dynamic exchange. Because this framework uses mruby, unlike the proposed memory-safe framework, memory safety must be guaranteed by the programmer.\par
    
    An emphasis framework for communication between Android and embedded devices is implemented by TECS. This framework enables Android smartphones to communicate with embedded devices. By implementing TECS plugins, the components necessary for communication are automatically generated. The components are not memory safety since they are developed for Android and in C. The proposed framework is for memory-safe CBD, which makes a difference.

\section{Conclusion}

In this paper, two TECS Rust plugins were proposed to realize memory-safe CBD for embedded systems. To create the Rust plugins, the Rust code generated from the component description files was determined. Then, a way to actually use these Rust plugins was shown. Not only a normal plugin, but also a plugin that enables easy use on the RTOS was proposed. The RTOS plugin used itron crate to simplify the use of the RTOS functionality from Rust. The code reusability was demonstrated by showing the number of lines of the code created using the Rust plugin. In addition, low overhead was demonstrated by comparing the execution time of the code adapted to the proposed framework with normal code.\par
In the future, the proposed framework plans to support various functionality of TECS. In addition, Rust components in TECS should be able to interoperate with components in other languages. Interoperability allows existing components from the proposed framework. This makes possible the use of rich resources such as C.

\vspace{7pt}

\bibliographystyle{IEEEtran}
\bibliography{bibtex}

\end{document}